\documentclass[a4paper,fleqn,usenatbib,onecolumn]{mnras}

\usepackage[T1]{fontenc}
\usepackage{ae,aecompl}

\usepackage{graphicx}	
\usepackage{amsmath}	
\usepackage{amssymb}	
\usepackage{multirow,bigstrut}
\usepackage{geometry}
\usepackage{slashbox}

\title[On the effect of pulsar evaporation on the cooling of white dwarfs]{On the effect of pulsar evaporation on the cooling of white dwarfs}
\author[Tang \& Li]
{Wenshi Tang$^{\rm 1}$, Xiang-Dong Li$^{\rm 1, 2}$\thanks{E-mail: lixd@nju.edu.cn;}\\
$^1$School of Astronomy and Space Science, Nanjing University, Nanjing 210023, China\\
$^2$ Key Laboratory of Modern Astronomy and Astrophysics (Nanjing University), Ministry of Education, Nanjing 210023, China}

\date{Accepted XXX. Received YYY; in original form ZZZ}

\pubyear{2020}

\begin{document}
\label{firstpage}
\pagerange{\pageref{firstpage}--\pageref{lastpage}}
\maketitle

\begin{abstract}\label{0. abstract}

Evolution of a large part of low-mass X-ray binaries (LMXBs) leads to the formation of rapidly rotating pulsars with a helium white dwarf (He WD) companion. Observations indicate that some He WDs in binary pulsar systems  are ultracool (with the effective temperatures $T_{\rm eff}\lesssim$ 4000\, K). It is hard to cool down a He WD to such low temperatures within the Hubble time, because a thick hydrogen envelope was left behind around the He core after the mass transfer process. A possible mechanism that can accelerate the WD cooling is the evaporative wind mass loss from the He WD driven by the high-energy radiation from the recycled pulsar. In this paper, we evolve a large number of LMXBs and investigate the influence of the pulsar's high-energy radiation on the WD cooling with different input parameters, including  the neutron star's spin-down luminosity, the evaporation efficiency and the metallicity of the companion star. By comparing our results with observations we note that, for relatively hot He WDs (with $T_{\rm eff}> 7000$ K), standard WD cooling without evaporation considered is able to reproduce their temperatures, while evaporation is probably required for the He WDs with relatively low temperatures ($T_{\rm eff}$ <5000 K).
\end{abstract}


\section{Introduction} \label{Introduction}
	Helium white dwarfs (He WDs) are the evolutionary products of low-mass stars.  A number of He WDs have been discovered  based on the survey observations such as the ELM Survey (Gianninas et al. \citeyear{2015ApJ...812..167G}; Brown et al. \citeyear{2016ApJ...818..155B}), Wide Angle Search for Planets (Maxted et al. \citeyear{2011MNRAS.418.1156M}), Palomar Transient Factory (van Roestel et al. \citeyear{2018MNRAS.475.2560V}), Kepler (van Kerkwijk et al. \citeyear{2010ApJ...715...51V}; Zhang et al. \citeyear{2017ApJ...850..125Z}), and Transiting Exoplanet Survey Satellite (Wang et al. \citeyear{2020ApJ...888...49W}). Shortly after their discovery, it was proposed that He WDs originate from binary evolution (Marsh et al. \citeyear{1995MNRAS.275..828M}). Some of the He WDs were found to have a pulsar companion, i.e. so-called low-mass binary pulsar systems (LMBPs; e.g. Camilo et al. \citeyear{2000ApJ...535..975C}; Bassa et al. \citeyear{2016MNRAS.455.3806B}). Currently, there are more than 140 known LMBPs in the Galaxy (e.g. Manchester et al. \citeyear{2005AJ....129.1993M}). The spin periods of the neutron stars (NSs) in LMBPs are of order of milliseconds (ms). It is widely accepted that LMBPs formed from low-mass X-ray binaries (LMXBs; e.g. Bhattacharya \& van den Heuvel \citeyear{1991PhR...203....1B}). In this scenario, a NS accretes mass and angular momentum from a Roche-lobe filling companion, and is spun up to a period of a few ms (Alpar \citeyear{1982Natur.300..728A}; Radhakrishnan \& Srinivasan \citeyear{1982CSci...51.1096R}). After the mass transfer, the core of the companion star evolves to be a hot He WD which cools with time. Very compact LMBPs will merge within Hubble time, being important  gravitational wave targets for upcoming space-based detectors such as Laser Interferometer Space Antenna (LISA; e.g. Nelemans \citeyear{2017arXiv170200786A}), Taiji (e.g. Ruan et al. \citeyear{2020IJMPA..3550075R} ), and TianQin observatories (e.g. Luo et al. \citeyear{2016CQGra..33c5010L}).

	An important subject in the study of LMBPs is cooling of the WD companions, which can provide independent information about the age of the systems. If we assume that the pulsar started to spin down by magnetic dipole radiation at the end of the mass-transfer, then the cooling age of the WD provides a unique opportunity to determine the pulsar's age, which is independent of the spin-down history of the pulsar. Models of WD cooling have been well established (Rohrmann et al. \citeyear{2002MNRAS.335..499R}; Holberg \& Bergeron \citeyear{2006AJ....132.1221H}; Kowalski \& Saumon \citeyear{2006ApJ...651L.137K}; Tremblay et al. \citeyear{2011ApJ...730..128T}; Bergeron et al. \citeyear{2011ApJ...737...28B}), and  used to compare with the measured temperatures of WDs in LMBPs. However, among the known He WDs in LMBPs, some have extremely low effective temperature  ($T_{\rm eff}\lesssim$ 4000\,K; e.g. Bassa et al. \citeyear{2016MNRAS.455.3806B}; Beronya et al. \citeyear{2019MNRAS.485.3715B}), and are classified as ultracool WDs (Gianninas et al. \citeyear{2015MNRAS.449.3966G}). They have posed challenge to the traditional theory of the cooling process in WDs.

	It is difficult to cool down a He WD to such a low temperature within Hubble time, because the hot He core is surrounded by a thick hydrogen envelope with a mass of order $10^{-2}\, M_{\rm \odot}$ (Sarna et al. \citeyear{2000MNRAS.316...84S}), and cooling of the WDs is slowed by the residual hydrogen burning. A possible mechanism that can  accelerate cooling is element diffusion (ED) caused by gravitational settling, thermal diffusion and chemical diffusion. Gravitational settling and thermal diffusion tend to transport heavier elements towards the center of the star, while chemical diffusion acts in the opposite direction (Thoul et al. \citeyear{1994ApJ...421..828T}). The effect of ED on the WD evolution has been studied by several authors (e.g. Althaus et al. \citeyear{2000MNRAS.317..952A}, \citeyear{2001MNRAS.323..471A}, \citeyear{2001MNRAS.324..617A}; Benvenuto \& Vito \citeyear{2004MNRAS.352..249B}, \citeyear{2005MNRAS.362..891B}; Istrate et al. \citeyear{2016A&A...595A..35I}). These investigations found that ED causes some hydrogen to move inwards to the hotter bottom layers, which may either induce another several hydrogen thermonuclear flashes or enhance the intensity of flashes. This can reduce the mass of the hydrogen in the envelope and make the WD to cool more quickly. For example, a decrease in the hydrogen envelope mass by a factor of 3 at the beginning of the cooling phase can reduce the cooling timescale of a $0.242\,M_{\rm \odot}$ WD  from 13 Gyr to 5 Gyr to reach $\log L/L_{\rm \odot}=-4$ (Althaus et al. \citeyear{2001ApJ...554.1110A}).

	However, the effect of ED relies on the occurrence of thermonuclear flashes. It is well known that there is a threshold mass ($M_{\rm th}$) for the occurrence of thermonuclear flashes, that is,  no flash occurs on the surface of WDs with a mass below $M_{\rm th}$. The value of $M_{\rm th}$ increases with decreasing metallicity (Serenelli et al. \citeyear{2002MNRAS.337.1091S}). For example, Istrate et al. (\citeyear{2016A&A...595A..35I}) found that $M_{\rm th}$ without and with ED considered is $\sim0.21, 0.22,0.25, 0.28\,M_{\rm \odot}$  and $\sim$ 0.167, 0.169, 0.21, 0.26 $M_{\rm \odot}$ for $Z=$ 0.02, 0.01, 0.001, 0.0002, respectively. The masses of some observed He WDs in LMBPs, such as PSRs J0613$-$0200 (Bassa et al.  \citeyear{2016MNRAS.455.3806B}) and J0740$+$6620 (Cromartie et al. \citeyear{2020NatAs...4...72C}), may be lower than $M_{\rm th}$. For example, PSR J0740+6620 is a LMBP in a 4.77 d orbit (Stovall et al. \citeyear{2014ApJ...791...67S}; Lynch et al. \citeyear{2018ApJ...859...93L}). Recently, Cromartie et al. (\citeyear{2020NatAs...4...72C}) measured the mass of PSR J0740+6620 to be $2.14^{\rm +0.20}_{\rm -0.18}\,M_{\rm \odot}$ and the mass of its He WD companion to be 0.26\,$^{\rm +0.016}_{\rm -0.014}\,M_{\rm \odot}$ (both at 95.4\% credibility), which may indicate that this system was born in a relatively low-metallicity environment (see also below). Recent optical and near-infrared observations revealed that the He WD is ultracool ($T_{\rm eff}$ < 3500 K; Beronya et al. \citeyear{2019MNRAS.485.3715B}). If the WD mass is indeed lower than $M_{\rm th}$, it would be difficult to account for the low temperature solely by the effect of ED.

	An alternative mechanism is evaporation of the companion by the pulsar's radiation. Once a millisecond pulsar (MSP) is formed, its high energy radiation and particles may evaporate its companion by heating and/or stripping the atmosphere (van den Heuvel \& van Paradijs \citeyear{1988Natur.334..227V}). This evaporation mechanism has been suggested to explain the formation of some specific systems such as redbacks, black widows and isolated MSPs (Chen et al. \citeyear{2013ApJ...775...27C}; Jia \& Li \citeyear{2016ApJ...830..153J}; Liu \& Li \citeyear{2017ApJ...851...58L}), as well as the rapid orbital expansion of PSR J0636+5128 (Chen et al. \citeyear{2021MNRAS.501.2327C}). In addition, this mechanism may also affect cooling of  the He WDs in LMBPs. Ergma et al. (\citeyear{2001MNRAS.321...71E})  proposed that evaporation of the companion can resolve the discrepancy between the spin-down age and the cooling age in PSR J0751$+$1807. Since Ergma et al. (\citeyear{2001MNRAS.321...71E}) focused on the specific system PSR J0751+1807, it is necessary to study the influence of evaporation with different parameters such as the initial orbital periods ($P_{\rm orb}^{\rm i}$), magnetic fields ($B$) and spin periods ($P_{\rm s}$) of the pulsars on the evolution of the WDs. Thus, this work aims to systematically investigate the effect of evaporation on the WD cooling and to explore to what extent it can account for the observed temperature of He WDs in LMBPs. 

	This paper is organized as follows.  In Section \ref{methods} we introduce the methods for calculations. The simulated results are demonstrated and compared with observations in Section \ref{results}. The summary is given in Section \ref{Discussion and Conclusions}.

\section{METHODS}\label{methods}
\subsection{Model}
	To produce a He WD with single star evolution requires a time longer than Hubble time, so it is widely accepted that He WDs result from binary interaction through stable mass-transfer or common-envelope evolution. The short spin periods and low magnetic fileds of the NSs in LMBPs imply that LMBPs likely formed through the former process \citep{1991PhR...203....1B}.

	There is mounting evidence that mass transfer in LMXBs is likely non-conservative (e.g. Jacoby et al. \citeyear{2005ApJ...629L.113J}; Antoniadis et al. \citeyear{2012MNRAS.423.3316A}). So we construct a model for the evolution of LMXBs assuming that 30 percent of the transferred matter from the low-mass donor star is accreted by the NS, and the remaining 70 percent of the material is ejected from the vicinity of the NS in the form of isotropic winds, taking away the specific angular momentum of the NS. Meanwhile, the mass accretion rate of the NS is also limited by the Eddington accretion rate. Moreover, angular momentum loss caused by magnetic braking and gravitational wave radiation is included (see Paxtion et al. \citeyear{2015ApJS..220...15P} for details).

	When the companion star detaches from its Roche-lobe and the mass transfer terminates, we assume that the NS behaves as a MSP, and evaporation commences (van den Heuvel \& van Paradijs \citeyear{1988Natur.334..227V}; Ruderman et al. \citeyear{1989ApJ...336..507R}). The mass loss rate ($\dot M_{\rm 2,evap}$) of the companion star caused by evaporation is given as follows ( van den Heuvel \& van Paradijs \citeyear{1988Natur.334..227V}; Stevens et al. \citeyear{1992MNRAS.254P..19S}):
\begin{equation} \label{E1}
\dot M_{\rm 2,evap}=-\frac{f}{2v_{\rm 2, esc}^{\rm 2}}L_{\rm p}\left({\frac{R_{\rm 2}}{a}}\right)^{\rm 2},
\end{equation}
where $f(<1)$ measures the fraction of the incident radiation energy that is invested in overcoming the companion's gravity. In principle it should be calculated by solving the wind's structure considering the interplay between thermal and dynamical processes, although it is usually left as a free parameter. In addition, $v_{\rm 2, esc}$ is the escape velocity at the surface of the companion star, $R_{2}$ is the radius of the companion star, and $a$ is the orbital separation of the system. The spin-down luminosity of the radio pulsar is $L_{\rm p}=4 \pi^{2}I\dot P_{\rm s}/P_{\rm s}^{3}$, where $I$ (taken to be $10^{45} \rm g\, cm^{2}$) is the moment of inertia of the pulsar, $P_{\rm s}$ the spin period of the pulsar and $\dot P_{\rm s}$ the derivative of the  spin period (e.g. Liu \& Li \citeyear{2017ApJ...851...58L}):
\begin{equation}\label{E2}
\dot P_{\rm s}=10^{-39}\frac{B^{2}}{P_{\rm s}},
\end{equation}
with $B$ being the surface magnetic field of the pulsar. Substitute Eq. (\ref{E2}) into Eq. (\ref{E1}), we obtain
\begin{equation}\label{E3}
\dot{M}_{\rm 2,evap}=-10^{-11}\Re\left(\frac{2\pi^{2}I}{v_{\rm 2, esc}^{\rm 2}}\right)\left({\frac{R_{\rm 2}}{a}}\right)^{\rm 2},
\end{equation}
with
\begin{equation}\label{Re}
\Re=f\frac{B_{8}^{2}}{P_{\rm s,ms}^{4}},
\end{equation}
which combines the evaporation efficiency and the pulsar's parameters into a single parameter. Here $B_{8}$ and $P_{\rm s,ms}$ are the magnetic field in units of $10^{8}$ G and the spin period in units of ms, respectively. We stop the evaporation process either when the envelope of the He core has been completely stripped or a Hubble time is reached. Meanwhile, we neglect any accretion onto the NS during the evaporation phase, and we will give a short discussion about this assumption in Sect.~\ref{Discussion and Conclusions}. 

\subsection{Numerical code and input parameters}
	The evolutionary tracks presented in this paper are calculated using the stellar evolution code Modules for Experiments in Stellar Astrophysics (MESA, version 10398; see Paxton et al. \citeyear{2011ApJS..192....3P}, \citeyear{2013ApJS..208....4P}, \citeyear{2015ApJS..220...15P}, \citeyear{2018ApJS..234...34P}). Cooling of WDs depends on the equations of state (EOS) and opacity of the atmosphere of the WDs. In MESA,  different EOS and opacity tables are used on different regions in the density-temperature ($\rho-T$) diagram. In the region of $2.5 \lesssim \log T (\rm K)\lesssim 7.5$ and $-10\lesssim\log \rho (\rm g/cm^{3})\lesssim 2$, OPAL EOS tables and SCVH tables, named as MESA EOS tables, are employed (Rogers \& Nayfonov \citeyear{2002ApJ...576.1064R}; Saumon et al. \citeyear{1995ApJS...99..713S}). Outside the MESA EOS region, HELM tables (Timmes \& Swesty \citeyear{2000ApJS..126..501T}) and PC tables (Potekhin \& Chabrier \citeyear{2010CoPP...50...82P}) are used (see Fig.~1 of Paxton et al. \citeyear{2011ApJS..192....3P}). The opacity tables are constructed from the work by Cassisi et al. (\citeyear{2007ApJ...661.1094C}), which combines radiation opacities with the electron conduction opacities. Radiation opacities are taken from Ferguson et al. (\citeyear{2005ApJ...623..585F}) for $2.7 \leq {\rm log} T (\rm K)\leq 4.5$, and OPAL (Iglesias \& Rogers \citeyear{1993ApJ...412..752I}, \citeyear{1996ApJ...464..943I}) for $3.75 \leq {\rm log} T (\rm K)\leq 8.5$. The energy-loss rates and their derivatives from neutrinos are taken from Itoh et al. (\citeyear{1996ApJS..102..411I}). In addition, convection is treated by standard mixing length theory by default (Cox \& Giuli \citeyear{1968pss..book.....C}).
	
In our calculations, the initial  masses of the neutron stars ($M_{\rm NS}^{\rm i}$) and the companions ($M_{\rm 2}^{\rm i}$) are set to be 1.4 $M_{\rm \odot}$ and 1.3 $M_{\rm \odot}$, respectively. We take the initial orbital period $P_{\rm orb}^{\rm i}$ to be longer than the bifurcation periods to guarantee the formation of He WDs for the evolution of LMXBs (Pylyser \& Savonije \citeyear{1988A&A...191...57P}, \citeyear{1989A&A...208...52P}). The ratio $\alpha$ of the mixing length to the pressure scale height is set to be 2.0 (e.g. Shao \& Li \citeyear{2012ApJ...756...85S}; Wang et al.  \citeyear{2014MNRAS.445.2340W}).

We calculate the binary evolution with three metallicities: $Z=0.02\, ({\rm Pop.\,I}), \,0.001 ({\rm Pop.\,II}),\, {\rm and}\, 0.0001$ (\rm Pop. III) considering the fact that some LMBPs may originate from a relatively low metallicity environment (Rivera-Sandoval et al. \citeyear{2015MNRAS.453.2707R}; Cadelano et al. \citeyear{2015ApJ...812...63C}; Cromartie et al. \citeyear{2020NatAs...4...72C}). 

\begin{figure*}
\centering
\includegraphics[width=0.5\textwidth]{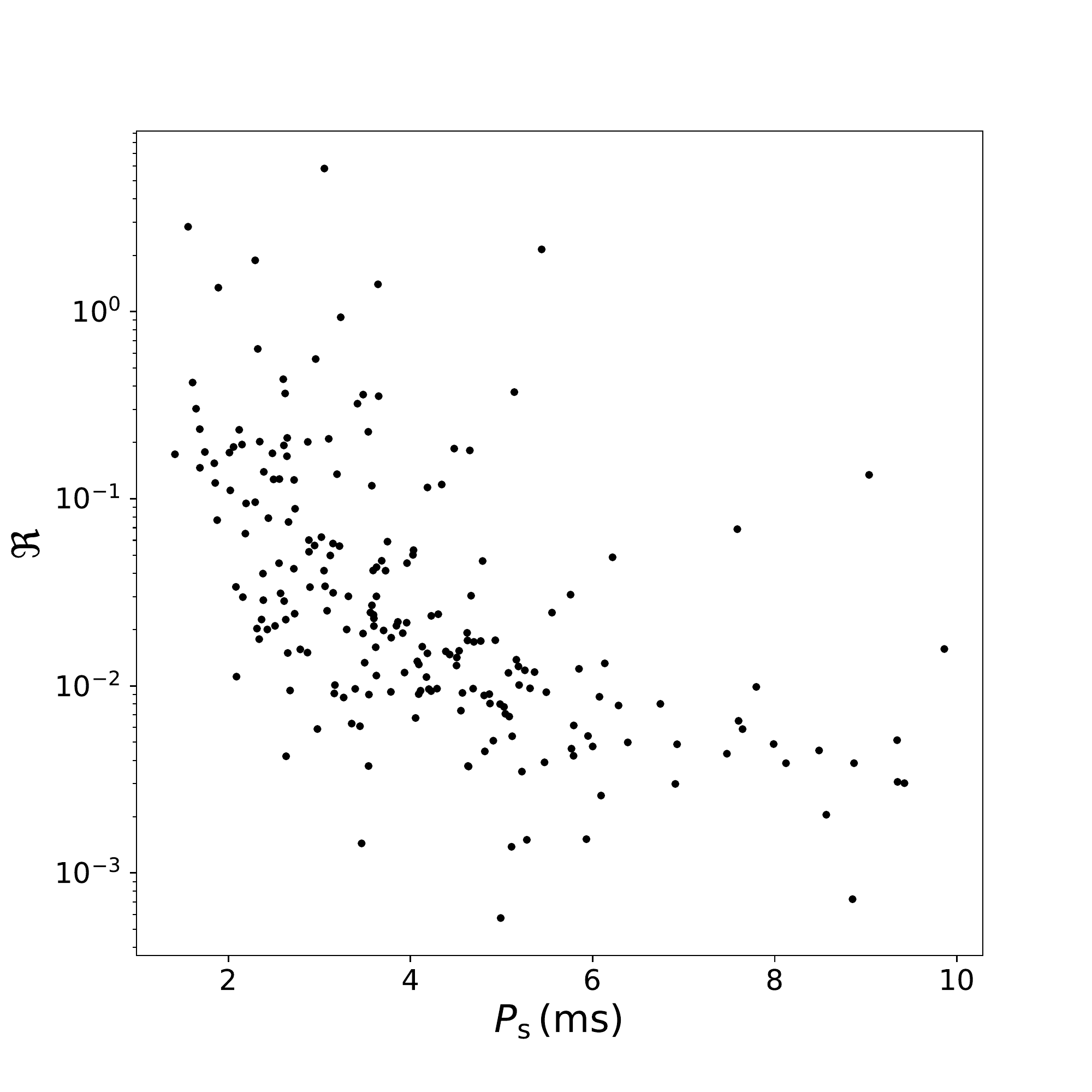}
 \caption{The $P_{\rm s}-\Re$ diagram for the observed MSPs with $P_{\rm s}\lesssim 10\rm ms$. The data are taken from the ATNF Pulsar Catalogue (Manchester et al. \citeyear{2005AJ....129.1993M}).}
 \label{Fig1}
\end{figure*}

One of the most important parameters in our work is $\Re$, because it determines the intensity of the evaporation power for a specific binary system. We constrain the possible range of $\Re$ from the observed parameters of radio pulsars. The $P_{\rm s}-\Re$ diagram is plotted in Fig. \ref{Fig1} for currently observed MSPs with $P_{\rm s}\le 10\,\rm ms$,  with $f$ set to be 1. We can roughly estimate the mean value  $\overline \Re \sim 0.137$ and the maximum value $\Re _{\rm max}\sim 6$.  In our calculations we adopt four values for $\Re=0$, 0.037, 0.37, and 1.0.

\section{results}\label{results}

\subsection{The $P_{\rm orb}-M_{\rm WD}$ relation}

\begin{figure*}
\centering
\includegraphics[width=0.3\textwidth]{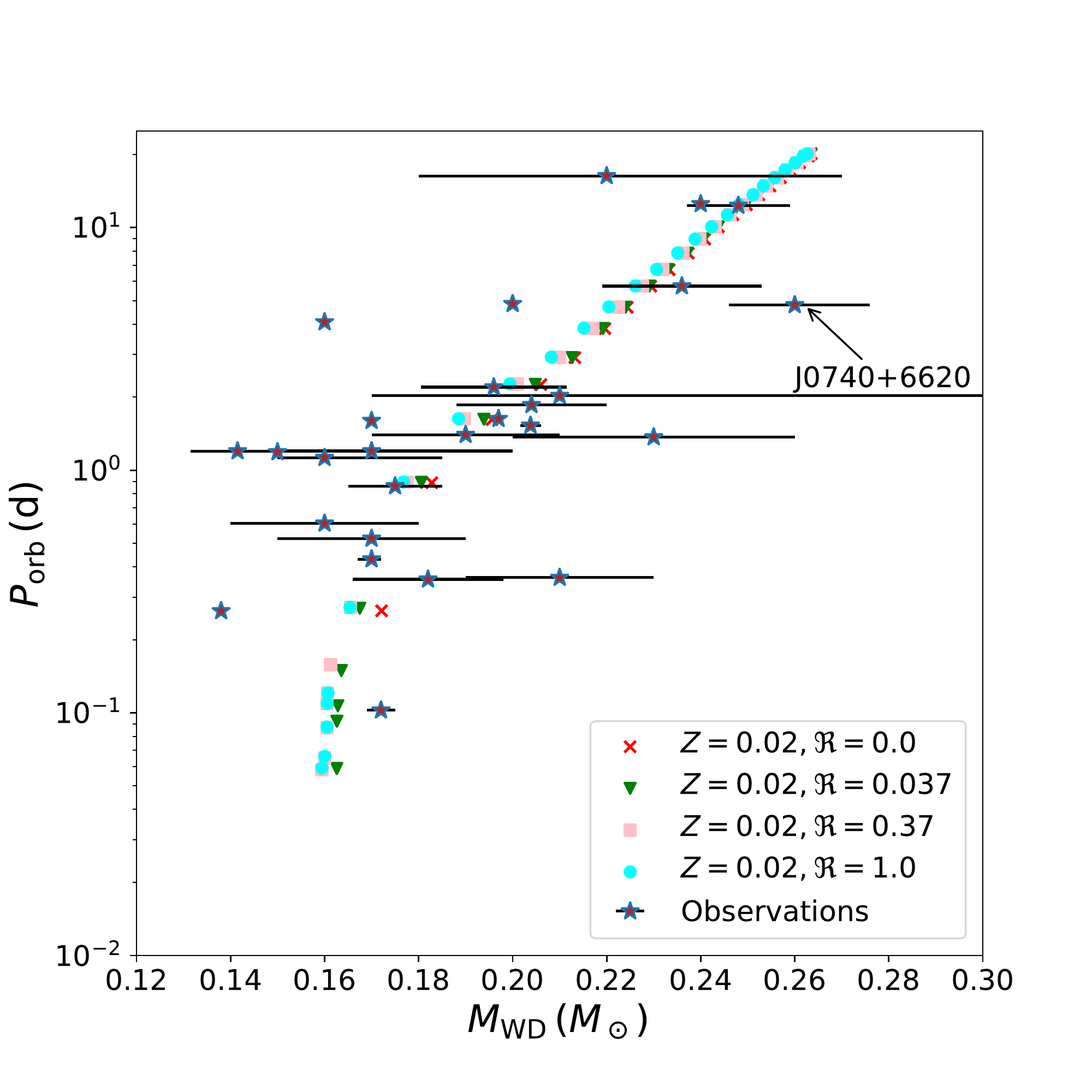}
\includegraphics[width=0.3\textwidth]{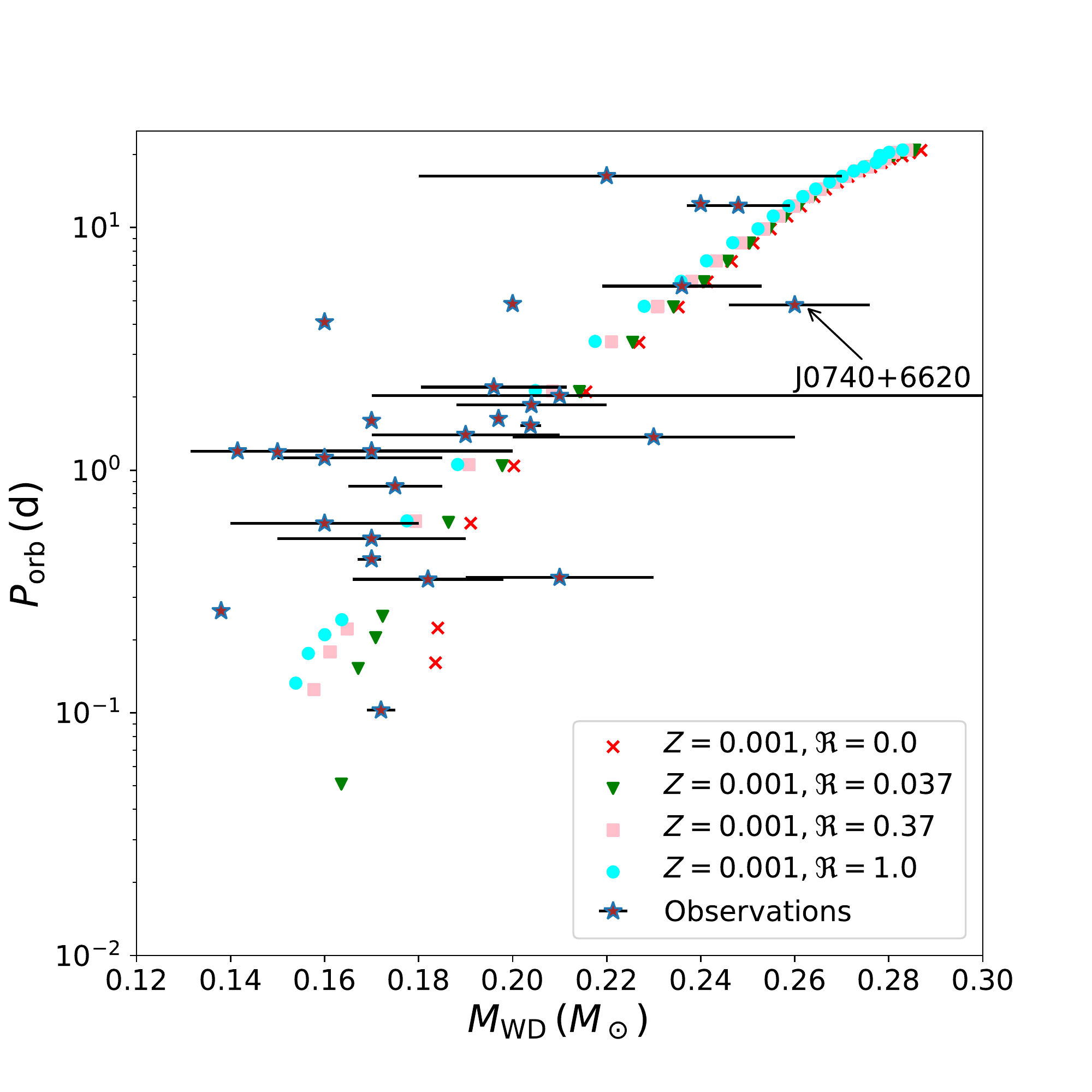}
\includegraphics[width=0.3\textwidth]{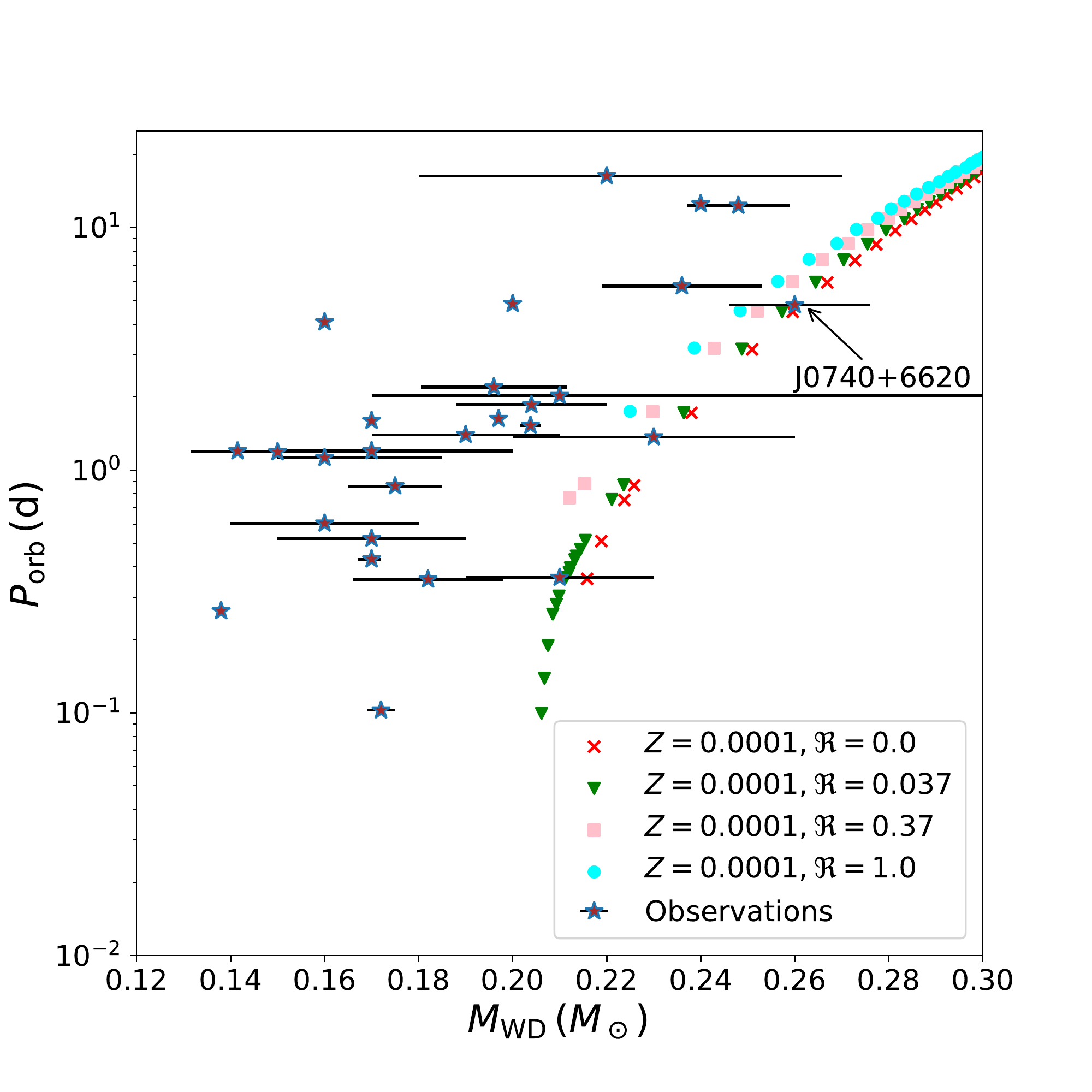}
 \caption{The final $P_{\rm orb}$-$M_{\rm WD}$ diagram. The calculated relations are shown as dots with different colors representing different values of $\Re$ (see the legends in figures).The stars with errorbars are observed LMBPs in which the companions' temperatures are available. The panels from left to right correspond to $Z=0.02,\,\,0.001, {\rm and} \,\,0.0001$, respectively.}
 \label{MWD-Porb}
\end{figure*}

	It is widely known that  there is a $P_{\rm orb}\,-\,M_{\rm WD}$ relation for the final outcomes of LMXB evolution (e.g. Tauris \& Savonije \citeyear{1999A&A...350..928T}; Lin et al. \citeyear{2011ApJ...732...70L}). Our simulated $P_{\rm orb}\,-\,M_{\rm WD}$ relations with different values of $\Re$ and $Z$ are shown in Fig.\,\ref{MWD-Porb}, and the LMBPs in which the He WD temperatures have been derived from observation are plotted with stars. Most of the observed sources follow the relations for Pops. I and II metallicities. However, we note that three sources, PSRs J0740+6620, J1641+3627F and J1816+4510,  seem to follow the relation of Pop. III, implying that they may originate from extremely low metallicity stars. In addition, the  $P_{\rm orb}-M_{\rm WD}$ relation becomes slightly scattered when different values of $\Re$ are adopted, especially for systems with short orbital periods, reflecting the decrease in the WD mass (or the increase in the orbital period) due to evaporation-induced mass loss. Therefore, the evaporation process can considerably alter the evolutionary tracks of the He WDs.

\subsection{The properties of the cooling WDs}\label{section3p3}
\begin{figure*}
\centering
\includegraphics[width=0.3\textwidth]{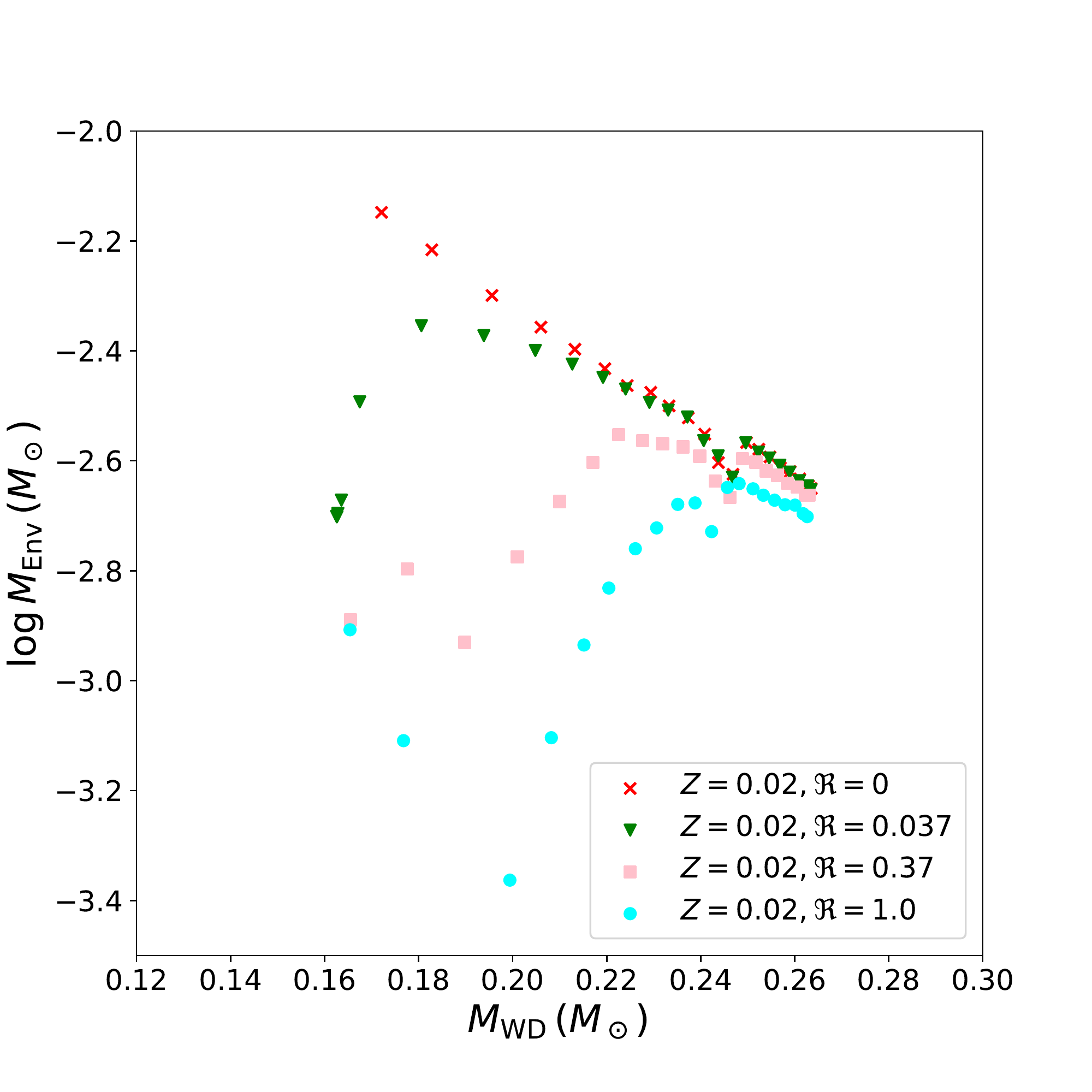}
\includegraphics[width=0.3\textwidth]{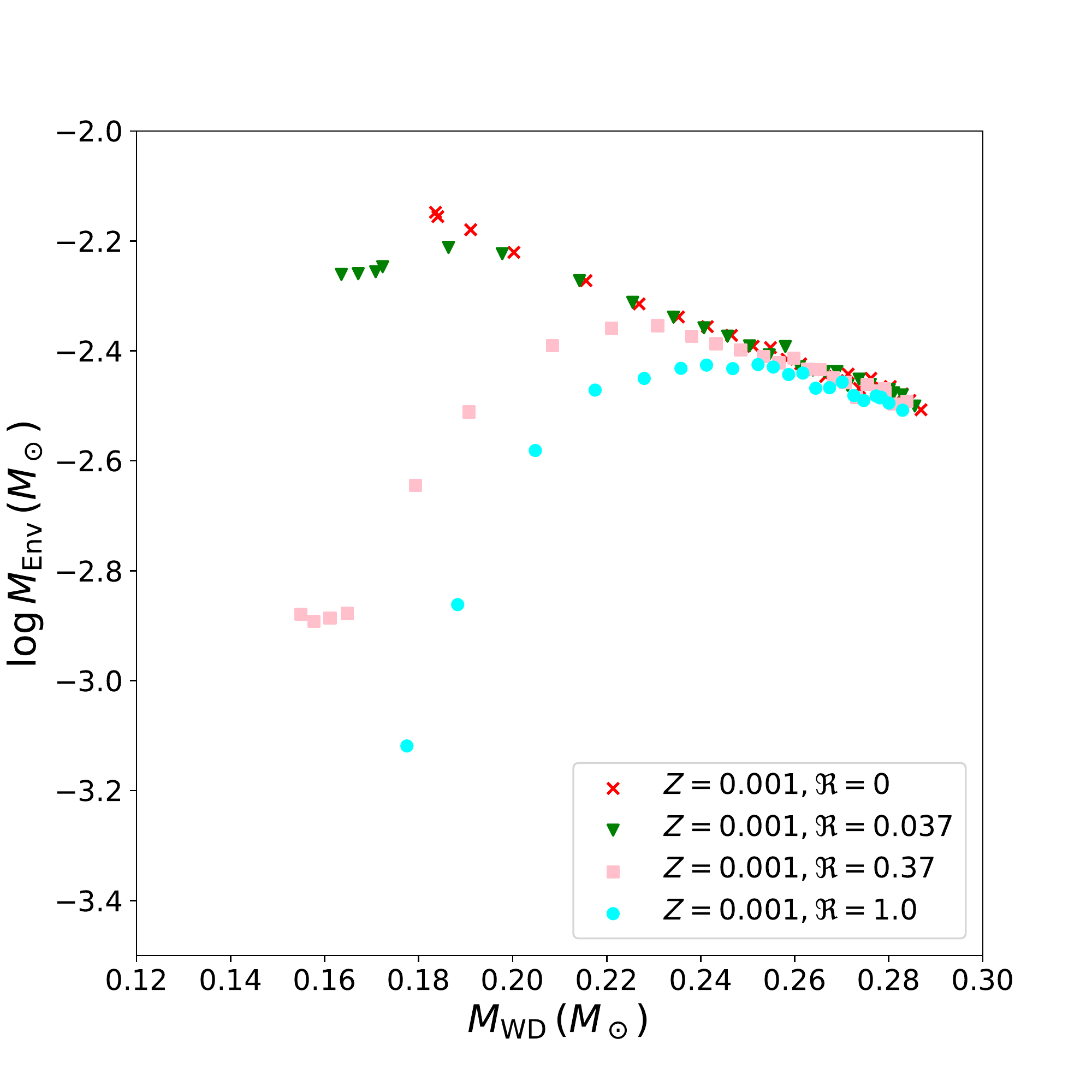}
\includegraphics[width=0.3\textwidth]{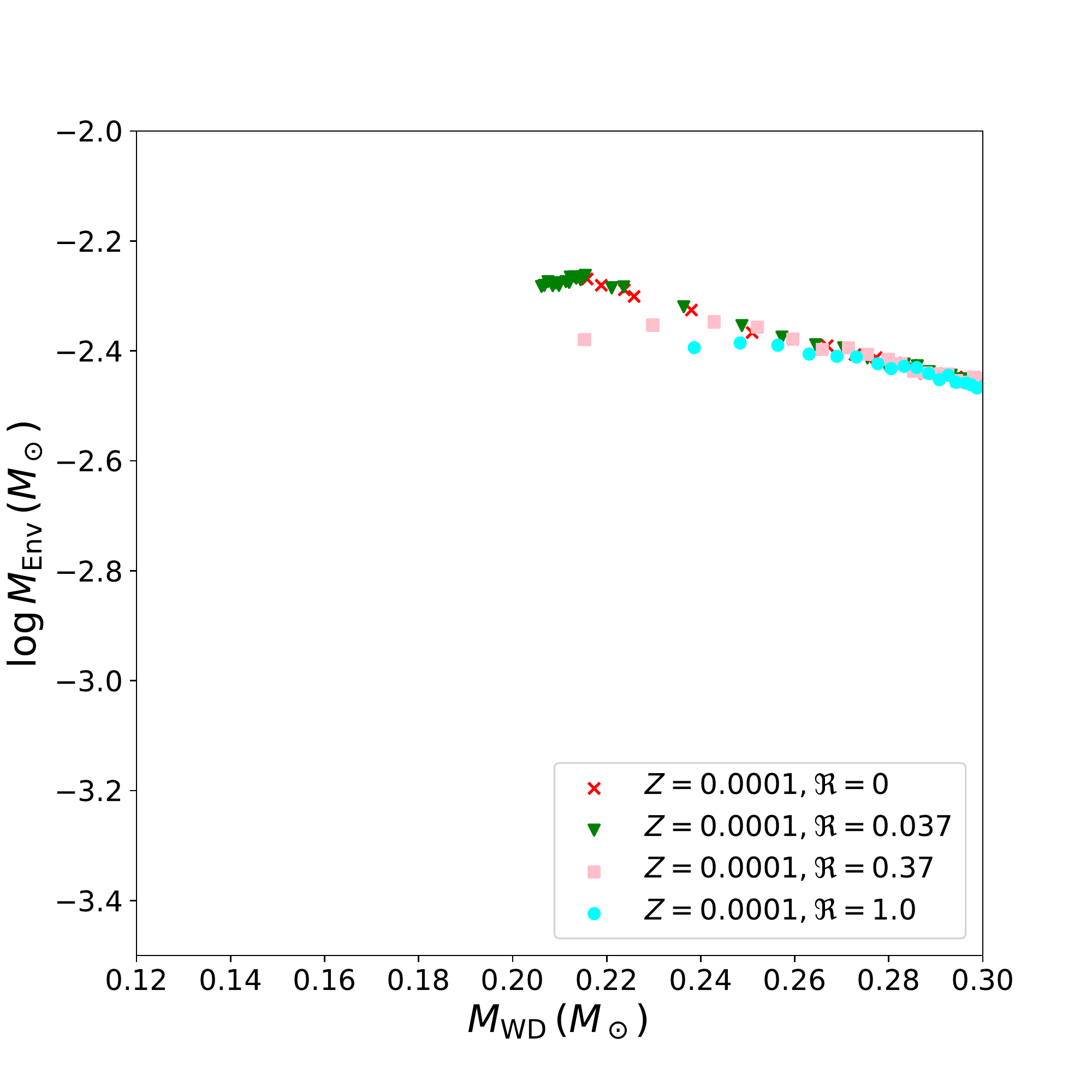}
 \caption{The final envelope mass $M_{\rm Env}$ as a function of the WD mass $M_{\rm WD}$. The panels from left to right are models with $Z=0.02,\,\,0.001, {\rm and}\,\,0.0001$, respectively}
 \label{MEnv}
\end{figure*}

	In Fig.\,\ref{MEnv}, the final mass ($M_{\rm Env}$) of the envelope on the He WDs is shown as a function of the final WD mass. The panels from left to right show the cases of $Z=0.02,\,\,0.001$, and $0.0001$, respectively. Beyond the left boundaries the calculation becomes non-convergent because of hitting the $\log Q$-limit (except for the case of $\Re=0$)\footnote{$\log Q = \log\rho-2\log T+12$, where $Q$ and $T$ are the density and temperature of a specific zone in the star model, respectively.}. The reason for the non-convergence may be that strong evaporation dramatically changes the structure of the WD's atmosphere.
Fig.\,\ref{MEnv} shows the following features:

 (1) The values of $M_{\rm Env}$ generally decrease with increasing $\Re$ for the same $Z$. This is because the mass loss rate is higher for larger value of $\Re$. However, this tendency becomes less significant with decreasing $Z$.

(2) There is more mass lost in narrower orbits (and with less massive WDs), but the amount of mass loss does not change monotonically with $M_{\rm WD}$ for the models with different values of $\Re$  (see below for the explanation).

	The values of $M_{\rm Env}$ decrease monotonically  with increasing WD mass in the case of $\Re =0$. This can be understood as follows. If there is no evaporation, the factors that can affect the final hydrogen content in the envelope are the initial hydrogen content in the proto-WDs and the residual hydrogen burning process. The amount of the residual hydrogen envelope at the end of the mass-transfer decreases with increasing proto-WD mass (see Fig.\,10 of Istrate et al. \citeyear{2016A&A...595A..35I}). Moreover, more massive WDs (with masses larger than $M_{\rm th}$) experience more flash cycles, which further consume the residual hydrogen (Althus et al. \citeyear{2000MNRAS.317..952A}, \citeyear{2001MNRAS.323..471A}, \citeyear{2001MNRAS.324..617A}; Benvenuto \& Vito \citeyear{2004MNRAS.352..249B}, \citeyear{2005MNRAS.362..891B}; Istrate et al. \citeyear{2016A&A...595A..35I}). Thus, more massive WDs have a less massive envelope and cool faster compared with less massive ones.

	However, the amount of the decrease in $M_{\rm Env}$ does not change monotonically with $M_{\rm WD}$ for the models with $\Re=0.037$, 0.37, and 1.0. This is actually the result of the competition between the hydrogen burning at the bottom layer of the envelope and mass loss from the outer layer stripped by evaporation. The pressure at the bottom layer of the envelope reduces due to mass loss, which results in insufficient hydrogen burning. The initially massive envelope combined with insufficient hydrogen burning may lead to a more massive envelope for less massive WDs. Beyond a threshold mass, the mass loss induced by evaporation becomes small, so evaporation can hardly affect the evolution of the envelope. The reasons are as follows. Firstly, a more massive WD has a smaller radius and  a larger escape velocity. Secondly, a more massive WD corresponds to a wider orbit. These factors give rise to a smaller wind mass loss rate induced by evaporation. Therefore, the evaporation process hardly affects the WD cooling in this case.

\begin{figure*}
\centering

\includegraphics[width=0.3\textwidth]{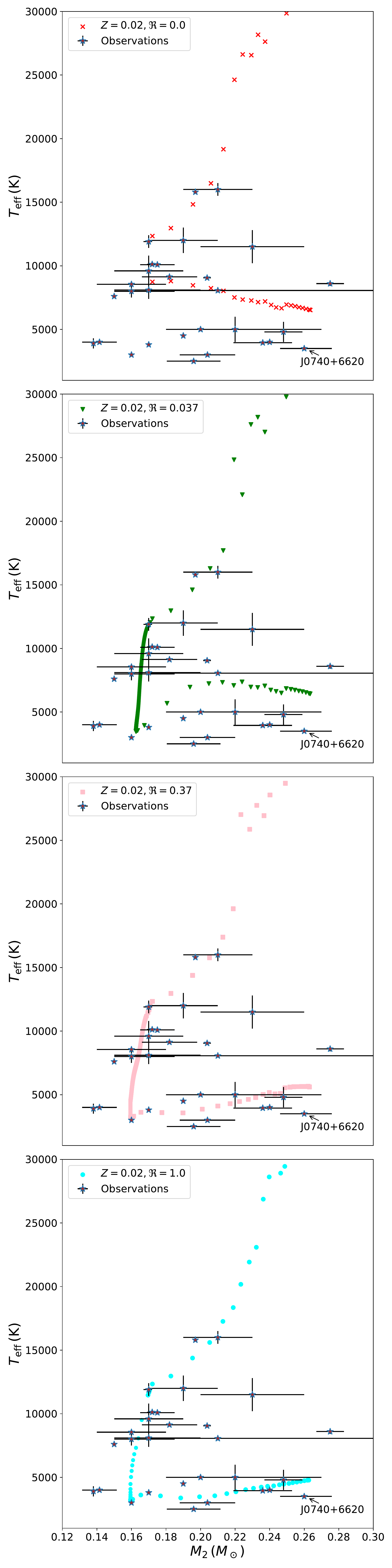}
\includegraphics[width=0.3\textwidth]{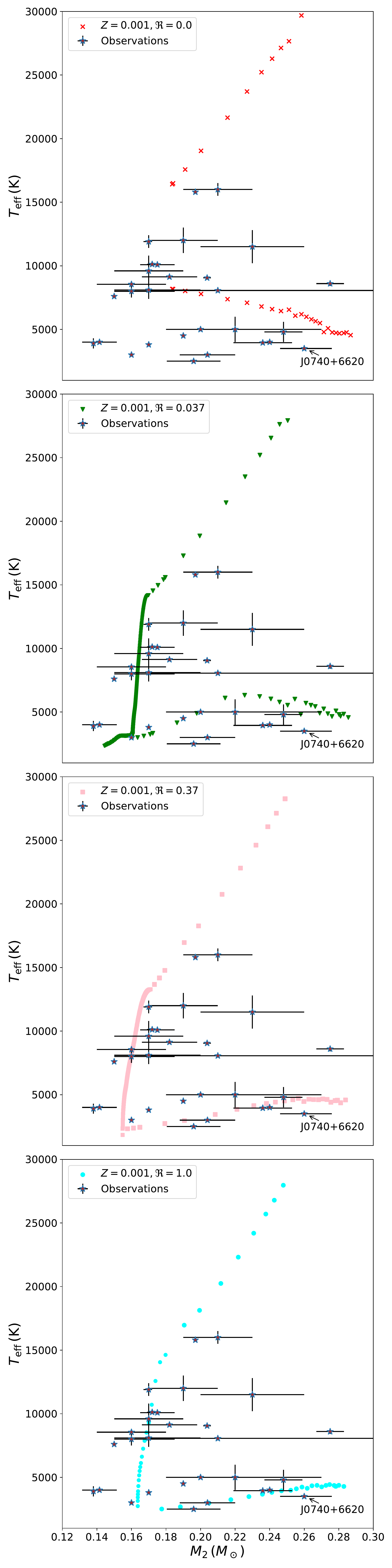}
\includegraphics[width=0.3\textwidth]{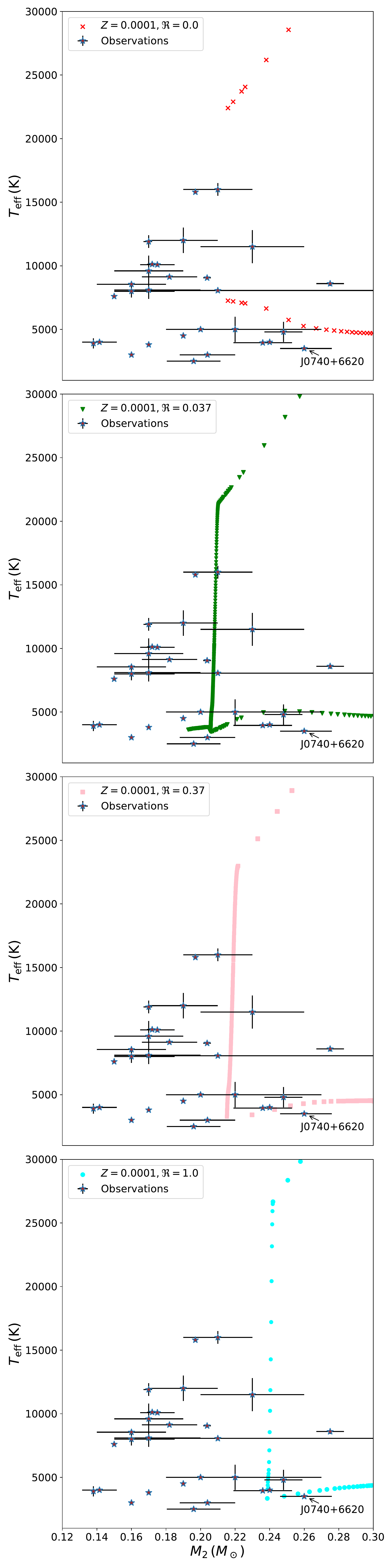}
 \caption{Distribution of the effective temperature of WDs as a function of the WD mass. The upper and lower boundaries represent the effective temperature at the beginning of the cooling phase and at Hubble time. The panels from top to bottom correspond to different adopted values of $\Re$. The panels from left to right correspond to the cases of $Z=0.02,\,\,0.001, {\rm and}\,\,0.0001$, respectively. The stars represent the observed LMBPs in which the companion temperatures are available. See Sect.\,\ref{comparison} for details. The data of the companion temperatures are from
 Antoniadis et al. (\citeyear{2012MNRAS.423.3316A});
 Antoniadis et al. (\citeyear{2013Sci...340..448A});
 Antoniadis (\citeyear{2013PhDT.......184A});
 Antoniadis et al. (\citeyear{2016ApJ...830...36A});
 Bassa et al. (\citeyear{2003A&A...403.1067B});
 Bassa et al. (\citeyear{2006A&A...456..295B});
 Bassa et al. (\citeyear{2006A&A...450..295B});
 Bassa et al. (\citeyear{2016MNRAS.455.3806B});
 Beronya et al. (\citeyear{2019MNRAS.485.3715B});
 Bobakov et al. (\citeyear{2019JPhCS1400b2023B});
 Cadelano et al. (\citeyear{2015ApJ...812...63C});
 Cadelano et al. (\citeyear{2019ApJ...875...25C});
 Cadelano et a. (\citeyear{2020ApJ...905...63C});
 Dai et al. (\citeyear{2017ApJ...842..105D});
 Durant et al. (\citeyear{2012ApJ...746....6D});
 Lundgren et al. (\citeyear{1996ASPC..105..497L});
 Kaplan et al. (\citeyear{2013ApJ...765..158K});
 Kaplan et al. (\citeyear{2014ApJ...783L..23K});
 Karpova et al. (\citeyear{2018PASA...35...28K});
 Kirichenko et al. (\citeyear{2020MNRAS.492.3032K});
 van Kerkwijk et al. (\citeyear{1996ApJ...467L..89V});
 van Kerkwijk et al. (\citeyear{2000ApJ...530L..37V});
 }
 \label{Obsevations}
\end{figure*}

	Fig.\,\ref{Obsevations} shows the distribution of the WDs in the $M_{\rm WD}-T_{\rm eff}$ diagram. Same as Fig.\,\ref{MEnv}, the panels from left to right show the cases with $Z=0.02$, 0.001, and 0.0001, respectively. The upper boundaries represent the effective temperature ($T_{\rm eff, max}$) at the beginning of the WD cooling process, and the lower boundaries represent the effective temperature at Hubble time. The closed left boundaries show the full cooling tracks.

	It can be seen from Fig\,\ref{Obsevations} that, for the models with the same $Z$, the WDs tend to cool down to lower $T_{\rm eff}$ with increasing $\Re$ for the same $M_{\rm WD}$. This is easy to understand because larger $\Re$ causes more envelope mass to be stripped and a smaller hydrogen envelope mass\footnote{The envelope is actually the H/He mixing atmosphere.} left behind, as shown in Fig.\,\ref{MEnv}. This clearly indicates that $M_{\rm Env}$ dramatically decreases with increasing $\Re$ for the same WD mass. For example, the ratio of $M_{\rm Env}$ in the cases of $\Re _{\rm max} (=1.0)$ and $\Re _{\rm min} (=0)$ can be as large as an order of magnitude for short  orbit period systems, which can lead to a significant difference (about several $10^{3}\,\rm K$) in $T_{\rm eff}$. In the cases of $\Re \neq 0$, the lowest value of $T_{\rm eff}$ decreases with increasing $\Re$, same as $M_{\rm Env}$, and the decrease in $T_{\rm eff}$ also depends on $M_{\rm WD}$.


	Furthermore, comparing the models with different $Z$ reveals two remarkable features. (1) The WDs with lower $Z$ can cool down to lower values of $T_{\rm eff}$ with the same $\Re$ and $M_{\rm WD}$. For example, in the case of $\Re=1$, a 0.26$M_{\rm \odot}$ WD can cool down to an effective temperature $\,\sim$ 4000\,K and 3000\,K for Pop. II and Pop. III metallicities respectively, but > 4000\,K for Pop. I metallicities. (2) From Figs.\,\ref{MEnv} and \ref{Obsevations}, it seems that smaller $M_{\rm Env}$ does not correspond to lower $T_{\rm eff}$ when we compare the models with different metallicities. Although the WDs with lower $Z$ have a slightly more massive envelope, their temperatures are lower than those with higher $Z$ but the same WD mass. For example, for a 0.26 $M_{\rm \odot}$ WD, the final values of $M_{\rm Env}$ increase with decreasing metallicity, i.e.,  about $10^{-3},\,10^{-2.6}$ and $10^{-2.5}\, M_{\rm \odot}$ for Pops. I, II and III metallicities,  respectively. However, the final values of $T_{\rm eff}$ decrease with decreasing metallicity.

	The reasons for the above two features are as follows. Firstly, from the $P_{\rm orb}-M_{\rm WD}$ relation (see Fig.\,\ref{MWD-Porb}), the orbital separation is smaller for lower metallicity for the same WD mass, which leads to a larger wind mass-loss rate. However, the envelope at the end of the mass transfer phase is more massive in the lower metallicity models (see Fig.\,10 of Istrate et al. \citeyear{2016A&A...595A..35I}). This explains why slightly more massive envelope is left behind in the lower metallicity models, though the systems have larger wind mass-loss rate. In addition, the opacity is smaller for lower metallicity. Thus, the WDs with lower metallicity cool faster than with higher metallicity.

\subsection{Comparison with observations}\label{comparison}
	We compare our results with the observed He WDs in LMBPs. In Fig.\,\ref{Obsevations}, we can see that for the case without evaporation (red dots), only part of the observed sources with relatively high temperatures can be accounted for by the calculated results. With increasing $\Re$, more observed sources, especially those with low temperatures, are covered by the calculated results. In  addition, there are several sources with extremely low temperatures which can only be explained by Pops. II or III models. We summarize that although the models without evaporation considered likely reproduce the masses and temperatures of WDs with $T_{\rm eff}>7000\,\rm K$, evaporation is probably required to explain those with $T_{\rm eff}\lesssim5000\,\rm K$.

\section{Discussion and Conclusions}\label{Discussion and Conclusions}
	In this work, we have evolved a large number of LMXB systems composed of a $1.4\,M_{\rm \odot}$ NS and a 1.3 $M_{\rm \odot}$ main sequence companion with different metallicities. We aim to investigate the effect of evaporation on the WD cooling by varying the evaporation efficiency $\Re$, the metallicity $Z$ and the initial orbital period $P_{\rm orb}^{\rm i}$. We analyze the relation between the final envelope mass, the effective temperature and the WD mass, and compare our results with the observations. Our main conclusions are as follows:

(1) Evaporation can considerably accelerate cooling of WDs.  Less massive envelope is left behind with increasing $\Re$ under the same initial condition. Therefore, WDs can evolve to lower temperatures when $\Re$ increases.

In addition, the final values of $T_{\rm eff}$ and $M_{\rm Env}$ depend on $M_{\rm WD}$ (or $P_{\rm orb}^{\rm i}$). The evolutions of $T_{\rm eff}$ and $M_{\rm Env}$ as a function of $M_{\rm WD}$ are different for different values of $\Re$. This is mainly the result of the competition between the residual hydrogen burning and the evaporation process.

 (2) For models with different metallicities, the WDs with lower $Z$ tend to evolve to lower $T_{\rm eff}$, although they have slightly more massive envelope with the same WD mass. The possible explanation is the difference in the initial envelope mass and  the opacity for different metallicities.

 (3) We find that the observed sources can be divided into two subpopulations. For the sources with relatively high temperatures ($T_{\rm eff}>7000\,\rm K$), standard WD cooling without evaporation considered is able to reproduce their temperatures. However, the evaporation process probably plays an important role in the cooling of the sources with relatively low temperatures ($T_{\rm eff}\lesssim5000$ K). Especially, some ultra-cool WDs ($T_{\rm eff}\lesssim4000\,\rm K$) may require efficient evaporation. In addition, for PSR J0740+6620 we also need to invoke Pop. III metallicities.

In our calculations we ignore any accretion by the NSs from the evaporated wind material. To justify this assumption we need to examine whether the radiation pressure from the pulsar is strong enough to expel the wind material outside the radius of the light cylinder, $R_{\rm lc}=cP_{\rm s}/2\pi$ ($\simeq 4.8\times 10^7$ cm for a 10 ms pulsar. Here $c$ is the speed of light). By comparing the stopping radius $R_{\rm st}$ with $R_{\rm lc}$, one can judge whether the wind material can penetrate into the light cylinder of the NS and influence its spin evolution. The stopping radius $R_{\rm st}$ where the ram pressure of the external wind is balanced by the radiation pressure is given by (Frank et al. \citeyear{2002apa..book.....F})
\begin{equation}\label{RB}
R_{\rm st}=\left(\frac{L_{\rm p}}{4\pi c\rho_{\rm w}v_{\rm rel}^{2}}\right)^{1/2},
\end{equation}
where
\begin{equation}
v_{\rm rel}=\sqrt{v_{\rm orb}^{2}+v_{\rm 2, esc}^{2}}
\end{equation}
is the relative velocity, $v_{\rm orb}$ is the orbital velocity of the NS, and
\begin{eqnarray}
\rho_{\rm w} & = & \frac{\dot{M}_{2, \rm evap}}{4\pi a^2v_{\rm 2, esc}} \\ \nonumber
 & \simeq & 3\times 10^{-16} \left(\frac{\dot{M}_{2,  \rm evap}}{10^{18}\,{\rm gs}^{-1}}\right)
\left(\frac{a}{10^{12}\,{\rm cm}}\right)^{-2}
\left(\frac{v_{\rm 2, esc}}{10^{8}\,{\rm cms}^{-1}}\right)^{-1}\,{\rm gcm}^{-3}
\end{eqnarray}
is the wind density at the location of the NS. The transition spin period obtained by equating $R_{\rm st}=R_{\rm lc}$ is \citep{Z04}
\begin{equation}
P_{\rm tr}\simeq 0.13 \left(\frac{B}{10^{9}\,{\rm G}}\right)\left(\frac{\rho_{\rm w}}{10^{-16}\,{\rm gcm}^{-3}}\right)^{-1/6}
\left(\frac{v_{\rm rel}}{10^{8}\,{\rm cms}^{-1}}\right)^{-1/3}\,{\rm s}.
\end{equation}
For MSPs, $P_{\rm s}\ll P_{\rm tr}$. Thus we can safely assume that evaporated wind does not affect the spin evolution of the pulsars.

Finally, we note that our results do not cover very low mass WDs ($M_{\rm WD}<0.15\,M_{\rm \odot}$) since the calculations becomes non-convergent in this case. They are unlikely accounted for by the effect of ED because their masses are below $M_{\rm th}$, which implies no flashes on their surface. Therefore, the cooling properties of these sources deserve further investigation.

\section*{acknowledgments}
We are grateful to the referee for helpful comments. This work was supported by the National Key Research and Development Program of China (2016YFA0400803), the Natural Science Foundation of China under grant No. 11773015, 12041301 and Project U1838201 supported by NSFC and CAS.

\section*{dataavailability}
The data underlying this article will be shared on reasonable request to the corresponding author.



\begin{thebibliography}{plain}\label{thebibliography}

\bibitem[Alpar et al.(1982)]{1982Natur.300..728A} Alpar, M.~A., Cheng, A.~F., Ruderman, M.~A., et al.\ 1982, \nat, 300, 728. doi:10.1038/300728a0

\bibitem[Althaus \& Benvenuto(2000)]{2000MNRAS.317..952A} Althaus, L.~G. \& Benvenuto, O.~G.\ 2000, \mnras, 317, 952. doi:10.1046/j.1365-8711.2000.03825.x

\bibitem[Althaus et al.(2001)]{2001MNRAS.323..471A} Althaus, L.~G., Serenelli, A.~M., \& Benvenuto, O.~G.\ 2001, \mnras, 323, 471. doi:10.1046/j.1365-8711.2001.04227.x

\bibitem[Althaus et al.(2001)]{2001MNRAS.324..617A} Althaus, L.~G., Serenelli, A.~M., \& Benvenuto, O.~G.\ 2001, \mnras, 324, 617. doi:10.1046/j.1365-8711.2001.04324.x

\bibitem[Althaus et al.(2001)]{2001ApJ...554.1110A} Althaus, L.~G., Serenelli, A.~M., \& Benvenuto, O.~G.\ 2001, \apj, 554, 1110. doi:10.1086/321414

\bibitem[Amaro-Seoane et al.(2017)]{2017arXiv170200786A} Amaro-Seoane, P., Audley, H., Babak, S., et al.\ 2017, arXiv:1702.00786

\bibitem[Antoniadis(2013)]{2013PhDT.......184A} Antoniadis, J.~I.\ 2013, Ph.D. Thesis

\bibitem[Antoniadis et al.(2012)]{2012MNRAS.423.3316A} Antoniadis, J., van Kerkwijk, M.~H., Koester, D., et al.\ 2012, \mnras, 423, 3316. doi:10.1111/j.1365-2966.2012.21124.x

\bibitem[Antoniadis et al.(2013)]{2013Sci...340..448A} Antoniadis, J., Freire, P.~C.~C., Wex, N., et al.\ 2013, Science, 340, 448. doi:10.1126/science.1233232

\bibitem[Antoniadis et al.(2016)]{2016ApJ...830...36A} Antoniadis, J., Kaplan, D.~L., Stovall, K., et al.\ 2016, \apj, 830, 36. doi:10.3847/0004-637X/830/1/36

\bibitem[Bassa et al.(2003)]{2003A&A...403.1067B} Bassa, C.~G., van Kerkwijk, M.~H., \& Kulkarni, S.~R.\ 2003, \aap, 403, 1067. doi:10.1051/0004-6361:20030384

\bibitem[Bassa et al.(2006)]{2006A&A...450..295B} Bassa, C.~G., van Kerkwijk, M.~H., \& Kulkarni, S.~R.\ 2006, \aap, 450, 295. doi:10.1051/0004-6361:20054316

\bibitem[Bassa et al.(2006)]{2006A&A...456..295B} Bassa, C.~G., van Kerkwijk, M.~H., Koester, D., et al.\ 2006, \aap, 456, 295. doi:10.1051/0004-6361:20065181

\bibitem[Bassa et al.(2016)]{2016MNRAS.455.3806B} Bassa, C.~G., Antoniadis, J., Camilo, F., et al.\ 2016, \mnras, 455, 3806. doi:10.1093/mnras/stv2607

\bibitem[Benvenuto \& De Vito(2004)]{2004MNRAS.352..249B} Benvenuto, O.~G. \& De Vito, M.~A.\ 2004, \mnras, 352, 249. doi:10.1111/j.1365-2966.2004.07918.x

\bibitem[Benvenuto \& De Vito(2005)]{2005MNRAS.362..891B} Benvenuto, O.~G. \& De Vito, M.~A.\ 2005, \mnras, 362, 891. doi:10.1111/j.1365-2966.2005.09315.x

\bibitem[Bergeron et al.(2011)]{2011ApJ...737...28B} Bergeron, P., Wesemael, F., Dufour, P., et al.\ 2011, \apj, 737, 28. doi:10.1088/0004-637X/737/1/28

\bibitem[Beronya et al.(2019)]{2019MNRAS.485.3715B} Beronya, D.~M., Karpova, A.~V., Kirichenko, A.~Y., et al.\ 2019, \mnras, 485, 3715. doi:10.1093/mnras/stz607

\bibitem[Bhattacharya \& van den Heuvel(1991)]{1991PhR...203....1B} Bhattacharya, D. \& van den Heuvel, E.~P.~J.\ 1991, \physrep, 203, 1. doi:10.1016/0370-1573(91)90064-S

\bibitem[Bobakov et al.(2019)]{2019JPhCS1400b2023B} Bobakov, A.~V., Zyuzin, D.~A., \& Shibanov, Y.~A.\ 2019, Journal of Physics Conference Series, 1400, 022023. doi:10.1088/1742-6596/1400/2/022023

\bibitem[Brown et al.(2016)]{2016ApJ...818..155B} Brown, W.~R., Gianninas, A., Kilic, M., et al.\ 2016, \apj, 818, 155. doi:10.3847/0004-637X/818/2/155

\bibitem[Cadelano et al.(2015)]{2015ApJ...812...63C} Cadelano, M., Pallanca, C., Ferraro, F.~R., et al.\ 2015, \apj, 812, 63. doi:10.1088/0004-637X/812/1/63

\bibitem[Cadelano et al.(2019)]{2019ApJ...875...25C} Cadelano, M., Ferraro, F.~R., Istrate, A.~G., et al.\ 2019, \apj, 875, 25. doi:10.3847/1538-4357/ab0e6b

\bibitem[Cadelano et al.(2020)]{2020ApJ...905...63C} Cadelano, M., Chen, J., Pallanca, C., et al.\ 2020, \apj, 905, 63. doi:10.3847/1538-4357/abc345

\bibitem[Camilo et al.(2000)]{2000ApJ...535..975C} Camilo, F., Lorimer, D.~R., Freire, P., et al.\ 2000, \apj, 535, 975. doi:10.1086/308859

\bibitem[Cassisi et al.(2007)]{2007ApJ...661.1094C} Cassisi, S., Potekhin, A.~Y., Pietrinferni, A., et al.\ 2007, \apj, 661, 1094. doi:10.1086/516819

\bibitem[Chen (2021)]{2021MNRAS.501.2327C} Chen, W.-C.\ 2021, \mnras, 501, 2327. doi:10.1093/mnras/staa3701

\bibitem[Chen et al.(2013)]{2013ApJ...775...27C} Chen, H.-L., Chen, X., Tauris, T.~M., et al.\ 2013, \apj, 775, 27. doi:10.1088/0004-637X/775/1/27

\bibitem[Cox \& Giuli(1968)]{1968pss..book.....C} Cox, J.~P. \& Giuli, R.~T.\ 1968, Principles of stellar structure, by J.P. Cox and R. T. Giuli.  New York: Gordon and Breach, 1968

\bibitem[Cromartie et al.(2020)]{2020NatAs...4...72C} Cromartie, H.~T., Fonseca, E., Ransom, S.~M., et al.\ 2020, Nature Astronomy, 4, 72. doi:10.1038/s41550-019-0880-2

\bibitem[Dai et al.(2017)]{2017ApJ...842..105D} Dai, S., Smith, M.~C., Wang, S., et al.\ 2017, \apj, 842, 105. doi:10.3847/1538-4357/aa7209

\bibitem[Durant et al.(2012)]{2012ApJ...746....6D} Durant, M., Kargaltsev, O., Pavlov, G.~G., et al.\ 2012, \apj, 746, 6. doi:10.1088/0004-637X/746/1/6


\bibitem[Ergma et al.(2001)]{2001MNRAS.321...71E} Ergma, E., Sarna, M.~J., \& Ger{\v{s}}kevit{\v{s}}-Antipova, J.\ 2001, \mnras, 321, 71. doi:10.1046/j.1365-8711.2001.04000.x

\bibitem[Frank et al.(2002)]{2002apa..book.....F} Frank, J., King, A., \& Raine, D.~J.\ 2002, Accretion Power in Astrophysics, by Juhan Frank and Andrew King and Derek Raine, pp. 398. ISBN 0521620538. Cambridge, UK: Cambridge University Press, February 2002., 398

\bibitem[Ferguson et al.(2005)]{2005ApJ...623..585F} Ferguson, J.~W., Alexander, D.~R., Allard, F., et al.\ 2005, \apj, 623, 585. doi:10.1086/428642

\bibitem[Gianninas et al.(2015)]{2015MNRAS.449.3966G} Gianninas, A., Curd, B., Thorstensen, J.~R., et al.\ 2015, \mnras, 449, 3966. doi:10.1093/mnras/stv545

\bibitem[Gianninas et al.(2015)]{2015ApJ...812..167G} Gianninas, A., Kilic, M., Brown, W.~R., et al.\ 2015, \apj, 812, 167. doi:10.1088/0004-637X/812/2/167

\bibitem[Holberg \& Bergeron(2006)]{2006AJ....132.1221H} Holberg, J.~B. \& Bergeron, P.\ 2006, \aj, 132, 1221. doi:10.1086/505938

\bibitem[Iglesias \& Rogers(1993)]{1993ApJ...412..752I} Iglesias, C.~A. \& Rogers, F.~J.\ 1993, \apj, 412, 752. doi:10.1086/172958

\bibitem[Iglesias \& Rogers(1996)]{1996ApJ...464..943I} Iglesias, C.~A. \& Rogers, F.~J.\ 1996, \apj, 464, 943. doi:10.1086/177381

\bibitem[Istrate et al.(2016)]{2016A&A...595A..35I} Istrate, A.~G., Marchant, P., Tauris, T.~M., et al.\ 2016, \aap, 595, A35. doi:10.1051/0004-6361/201628874

\bibitem[Itoh et al.(1996)]{1996ApJS..102..411I} Itoh, N., Hayashi, H., Nishikawa, A., et al.\ 1996, \apjs, 102, 411. doi:10.1086/192264

\bibitem[Jacoby et al.(2005)]{2005ApJ...629L.113J} Jacoby, B.~A., Hotan, A., Bailes, M., et al.\ 2005, \apjl, 629, L113. doi:10.1086/449311

\bibitem[Jia \& Li(2016)]{2016ApJ...830..153J} Jia, K. \& Li, X.-D.\ 2016, \apj, 830, 153. doi:10.3847/0004-637X/830/2/153

\bibitem[Kaplan et al.(2013)]{2013ApJ...765..158K} Kaplan, D.~L., Bhalerao, V.~B., van Kerkwijk, M.~H., et al.\ 2013, \apj, 765, 158. doi:10.1088/0004-637X/765/2/158

\bibitem[Kaplan et al.(2014)]{2014ApJ...783L..23K} Kaplan, D.~L., van Kerkwijk, M.~H., Koester, D., et al.\ 2014, \apjl, 783, L23. doi:10.1088/2041-8205/783/1/L23

\bibitem[Karpova et al.(2018)]{2018PASA...35...28K} Karpova, A.~V., Zyuzin, D.~A., Shibanov, Y.~A., et al.\ 2018, \pasa, 35, e028. doi:10.1017/pasa.2018.21

\bibitem[Kirichenko et al.(2020)]{2020MNRAS.492.3032K} Kirichenko, A.~Y., Karpova, A.~V., Zyuzin, D.~A., et al.\ 2020, \mnras, 492, 3032. doi:10.1093/mnras/staa066

\bibitem[Kowalski \& Saumon(2006)]{2006ApJ...651L.137K} Kowalski, P.~M. \& Saumon, D.\ 2006, \apjl, 651, L137. doi:10.1086/509723

\bibitem[Lin et al.(2011)]{2011ApJ...732...70L} Lin, J., Rappaport, S., Podsiadlowski, P., et al.\ 2011, \apj, 732, 70. doi:10.1088/0004-637X/732/2/70

\bibitem[Liu \& Li(2017)]{2017ApJ...851...58L} Liu, W.-M. \& Li, X.-D.\ 2017, \apj, 851, 58. doi:10.3847/1538-4357/aa9922

\bibitem[Lundgren et al.(1996)]{1996ASPC..105..497L} Lundgren, S.~C., Foster, R.~S., \& Camilo, F.\ 1996, IAU Colloq. 160: Pulsars: Problems and Progress, 105, 497

\bibitem[Luo et al.(2016)]{2016CQGra..33c5010L} Luo, J., Chen, L.-S., Duan, H.-Z., et al.\ 2016, Classical and Quantum Gravity, 33, 035010. doi:10.1088/0264-9381/33/3/035010

\bibitem[Lynch et al.(2018)]{2018ApJ...859...93L} Lynch, R.~S., Swiggum, J.~K., Kondratiev, V.~I., et al.\ 2018, \apj, 859, 93. doi:10.3847/1538-4357/aabf8a

\bibitem[Manchester et al.(2005)]{2005AJ....129.1993M} Manchester, R.~N., Hobbs, G.~B., Teoh, A., et al.\ 2005, \aj, 129, 1993. doi:10.1086/428488

\bibitem[Marsh et al.(1995)]{1995MNRAS.275..828M} Marsh, T.~R., Dhillon, V.~S., \& Duck, S.~R.\ 1995, \mnras, 275, 828. doi:10.1093/mnras/275.3.828

\bibitem[Maxted et al.(2011)]{2011MNRAS.418.1156M} Maxted, P.~F.~L., Anderson, D.~R., Burleigh, M.~R., et al.\ 2011, \mnras, 418, 1156. doi:10.1111/j.1365-2966.2011.19567.x

\bibitem[Paxton et al.(2011)]{2011ApJS..192....3P} Paxton, B., Bildsten, L., Dotter, A., et al.\ 2011, \apjs, 192, 3. doi:10.1088/0067-0049/192/1/3

\bibitem[Paxton et al.(2013)]{2013ApJS..208....4P} Paxton, B., Cantiello, M., Arras, P., et al.\ 2013, \apjs, 208, 4. doi:10.1088/0067-0049/208/1/4

\bibitem[Paxton et al.(2015)]{2015ApJS..220...15P} Paxton, B., Marchant, P., Schwab, J., et al.\ 2015, \apjs, 220, 15. doi:10.1088/0067-0049/220/1/15

\bibitem[Paxton et al.(2018)]{2018ApJS..234...34P} Paxton, B., Schwab, J., Bauer, E.~B., et al.\ 2018, \apjs, 234, 34. doi:10.3847/1538-4365/aaa5a8

\bibitem[Perna et al.(2003)]{2003ApJ...594..936P} Perna, R., Narayan, R., Rybicki, G., et al.\ 2003, \apj, 594, 936. doi:10.1086/377091

\bibitem[Potekhin \& Chabrier(2010)]{2010CoPP...50...82P} Potekhin, A.~Y. \& Chabrier, G.\ 2010, Contributions to Plasma Physics, 50, 82. doi:10.1002/ctpp.201010017

\bibitem[Pylyser \& Savonije(1988)]{1988A&A...191...57P} Pylyser, E. \& Savonije, G.~J.\ 1988, \aap, 191, 57

\bibitem[Pylyser \& Savonije(1989)]{1989A&A...208...52P} Pylyser, E.~H.~P. \& Savonije, G.~J.\ 1989, \aap, 208, 52

\bibitem[Radhakrishnan \& Srinivasan(1982)]{1982CSci...51.1096R} Radhakrishnan, V. \& Srinivasan, G.\ 1982, Current Science, 51, 1096

\bibitem[Rivera-Sandoval et al.(2015)]{2015MNRAS.453.2707R} Rivera-Sandoval, L.~E., van den Berg, M., Heinke, C.~O., et al.\ 2015, \mnras, 453, 2707. doi:10.1093/mnras/stv1810

\bibitem[Rogers \& Nayfonov(2002)]{2002ApJ...576.1064R} Rogers, F.~J. \& Nayfonov, A.\ 2002, \apj, 576, 1064. doi:10.1086/341894

\bibitem[Rohrmann et al.(2002)]{2002MNRAS.335..499R} Rohrmann, R.~D., Serenelli, A.~M., Althaus, L.~G., et al.\ 2002, \mnras, 335, 499. doi:10.1046/j.1365-8711.2002.05644.x

\bibitem[Ruan et al.(2020)]{2020IJMPA..3550075R} Ruan, W.-H., Guo, Z.-K., Cai, R.-G., et al.\ 2020, International Journal of Modern Physics A, 35, 2050075. doi:10.1142/S0217751X2050075X

\bibitem[Ruderman et al.(1989)]{1989ApJ...336..507R} Ruderman, M., Shaham, J., \& Tavani, M.\ 1989, \apj, 336, 507. doi:10.1086/167029

\bibitem[Sarna et al.(2000)]{2000MNRAS.316...84S} Sarna, M.~J., Ergma, E., \& Ger{\v{s}}kevit{\v{s}}-Antipova, J.\ 2000, \mnras, 316, 84. doi:10.1046/j.1365-8711.2000.03503.x

\bibitem[Saumon et al.(1995)]{1995ApJS...99..713S} Saumon, D., Chabrier, G., \& van Horn, H.~M.\ 1995, \apjs, 99, 713. doi:10.1086/192204

\bibitem[Serenelli et al.(2002)]{2002MNRAS.337.1091S} Serenelli, A.~M., Althaus, L.~G., Rohrmann, R.~D., et al.\ 2002, \mnras, 337, 1091. doi:10.1046/j.1365-8711.2002.05994.x

\bibitem[Shao \& Li(2012)]{2012ApJ...756...85S} Shao, Y. \& Li, X.-D.\ 2012, \apj, 756, 85. doi:10.1088/0004-637X/756/1/85

\bibitem[Stevens et al.(1992)]{1992MNRAS.254P..19S} Stevens, I.~R., Rees, M.~J., \& Podsiadlowski, P.\ 1992, \mnras, 254, 19P. doi:10.1093/mnras/254.1.19P

\bibitem[Stovall et al.(2014)]{2014ApJ...791...67S} Stovall, K., Lynch, R.~S., Ransom, S.~M., et al.\ 2014, \apj, 791, 67. doi:10.1088/0004-637X/791/1/67

\bibitem[Tauris \& Savonije(1999)]{1999A&A...350..928T} Tauris, T.~M. \& Savonije, G.~J.\ 1999, \aap, 350, 928

\bibitem[Tauris et al.(2012)]{2012MNRAS.425.1601T} Tauris, T.~M., Langer, N., \& Kramer, M.\ 2012, \mnras, 425, 1601. doi:10.1111/j.1365-2966.2012.21446.x

\bibitem[Thoul et al.(1994)]{1994ApJ...421..828T} Thoul, A.~A., Bahcall, J.~N., \& Loeb, A.\ 1994, \apj, 421, 828. doi:10.1086/173695

\bibitem[Timmes \& Swesty(2000)]{2000ApJS..126..501T} Timmes, F.~X. \& Swesty, F.~D.\ 2000, \apjs, 126, 501. doi:10.1086/313304

\bibitem[Tremblay et al.(2011)]{2011ApJ...730..128T} Tremblay, P.-E., Bergeron, P., \& Gianninas, A.\ 2011, \apj, 730, 128. doi:10.1088/0004-637X/730/2/128

\bibitem[Wang et al.(2020)]{2020ApJ...888...49W} Wang, K., Zhang, X., \& Dai, M.\ 2020, \apj, 888, 49. doi:10.3847/1538-4357/ab584c

\bibitem[Wang et al.(2014)]{2014MNRAS.445.2340W} Wang, B., Justham, S., Liu, Z.-W., et al.\ 2014, \mnras, 445, 2340. doi:10.1093/mnras/stu1891

\bibitem[Zhang et al.(2017)]{2017ApJ...850..125Z} Zhang, X.~B., Fu, J.~N., Liu, N., et al.\ 2017, \apj, 850, 125. doi:10.3847/1538-4357/aa9577

\bibitem[van Kerkwijk et al.(1996)]{1996ApJ...467L..89V} van Kerkwijk, M.~H., Bergeron, P., \& Kulkarni, S.~R.\ 1996, \apjl, 467, L89. doi:10.1086/310209

\bibitem[van Kerkwijk et al.(2000)]{2000ApJ...530L..37V} van Kerkwijk, M.~H., Bell, J.~F., Kaspi, V.~M., et al.\ 2000, \apjl, 530, L37. doi:10.1086/312478

\bibitem[van Kerkwijk et al.(2010)]{2010ApJ...715...51V} van Kerkwijk, M.~H., Rappaport, S.~A., Breton, R.~P., et al.\ 2010, \apj, 715, 51. doi:10.1088/0004-637X/715/1/51

\bibitem[van Roestel et al.(2018)]{2018MNRAS.475.2560V} van Roestel, J., Kupfer, T., Ruiz-Carmona, R., et al.\ 2018, \mnras, 475, 2560. doi:10.1093/mnras/stx3291

\bibitem[van den Heuvel \& van Paradijs(1988)]{1988Natur.334..227V} van den Heuvel, E.~P.~J. \& van Paradijs, J.\ 1988, \nat, 334, 227. doi:10.1038/334227a0

\bibitem[Zhang, Li, \& Wang(2004)]{Z04} Zhang, F., Li, X.-D., \& Wang, Z.-R.  2004, RAA, 4, 320

\end{thebibliography}

\section*{dataavailability}
The data underlying this article will be shared on reasonable request to the corresponding author.

\bsp	
\label{lastpage}
\end{document}